\documentclass[twocolumn]{aastex63}

\usepackage{subfigure}
\usepackage{url}
\usepackage{hyperref}
\usepackage{epsfig}
\usepackage{graphicx}
\usepackage{sidecap}
\usepackage{hanging}
\usepackage{color}
\usepackage{multirow}
\usepackage{enumerate}
\usepackage{natbib}
\usepackage{amssymb}
\usepackage{amsmath}
\usepackage[bottom]{footmisc}

\newcommand{\sci}{Science}
\newcommand{\jatis}{JATIS}

\newcommand{\kepler}{\textit{Kepler}}
\newcommand{\tess}{{TESS}}

\newcommand{\gaia}{{Gaia}}

\providecommand{\e}[1]{\ensuremath{\, \times \, 10^{#1}}}

\newcommand{\koi}{KOI-3681.02}
\newcommand{\planet}{{Kepler-1514~b}}
\newcommand{\host}{{Kepler-1514}}

\providecommand{\bjdtdb}{\ensuremath{\rm {BJD_{TDB}}}}

\providecommand{\msun}{\ensuremath{\,M_\Sun}}
\providecommand{\rsun}{\ensuremath{\,R_\Sun}}
\providecommand{\lsun}{\ensuremath{\,L_\Sun}}
\providecommand{\mj}{\ensuremath{\,M_{\rm J}}}
\providecommand{\rj}{\ensuremath{\,R_{\rm J}}}

\providecommand{\fave}{\langle F \rangle}
\providecommand{\fluxcgs}{10$^9$ erg s$^{-1}$ cm$^{-2}$}
\shorttitle{GOT `EM I. A Dense, Cool Giant Planet Orbiting \host}
\shortauthors{Dalba et al.}

\begin{document}

\title{Giant Outer Transiting Exoplanet Mass (GOT `EM) Survey. I. Confirmation of an Eccentric, Cool Jupiter With an Interior Earth-sized Planet Orbiting Kepler-1514\footnote{Some of the data presented herein were obtained at the W. M. Keck Observatory, which is operated as a scientific partnership among the California Institute of Technology, the University of California and the National Aeronautics and Space Administration. The Observatory was made possible by the generous financial support of the W. M. Keck Foundation.}}

\correspondingauthor{Paul A. Dalba}
\email{pdalba@ucr.edu}

%%%%%%%%%%%%%%%%%%%%%%%%%%%%%%%%%%%%%%%%%%%%%%%%%%%%%%%%%%%%%%%%%%%%%%%%%%

\author[0000-0002-4297-5506]{Paul A.\ Dalba}
\altaffiliation{NSF Astronomy and Astrophysics Postdoctoral Fellow}
\affiliation{Department of Earth and Planetary Sciences, University of California Riverside, 900 University Ave, Riverside, CA 92521, USA}

\author[0000-0002-7084-0529]{Stephen R.\ Kane}
\affiliation{Department of Earth and Planetary Sciences, University of California Riverside, 900 University Ave, Riverside, CA 92521, USA}

\author[0000-0002-0531-1073]{Howard Isaacson}
\affiliation{{Department of Astronomy,  University of California Berkeley, Berkeley CA 94720, USA}}
\affiliation{Centre for Astrophysics, University of Southern Queensland, Toowoomba, QLD, Australia}

\author[0000-0002-8965-3969]{Steven Giacalone}
\affiliation{Department of Astronomy, University of California Berkeley, Berkeley, CA 94720-3411, USA}

\author[0000-0001-8638-0320]{Andrew W.\ Howard}
\affiliation{Department of Astronomy, California Institute of Technology, Pasadena, CA 91125, USA}

\author[0000-0001-8812-0565]{Joseph E.\ Rodriguez} 
\affiliation{Center for Astrophysics \textbar \ Harvard \& Smithsonian, 60 Garden St, Cambridge, MA 02138, USA}
\affiliation{Department of Physics and Astronomy, Michigan State University, East Lansing, MI 48824, USA}

\author[0000-0001-7246-5438]{Andrew Vanderburg}
\altaffiliation{NASA Sagan Fellow}
\affiliation{Department of Astronomy, University of Wisconsin-Madison, Madison, WI 53706, USA}
\affiliation{Department of Astronomy, The University of Texas at Austin, Austin, TX 78712, USA}

\author[0000-0003-3773-5142]{Jason D.\ Eastman}
\affiliation{Center for Astrophysics \textbar \ Harvard \& Smithsonian, 60 Garden St, Cambridge, MA 02138, USA}

\author[0000-0001-9811-568X]{Adam L.\ Kraus}
\affiliation{Department of Astronomy, The University of Texas at Austin, Austin, TX 78712, USA}

\author[0000-0001-9823-1445]{Trent J.\ Dupuy}
\affiliation{Institute for Astronomy, University of Edinburgh, Royal Observatory, Blackford Hill, Edinburgh, EH9 3HJ, UK}

\author[0000-0002-3725-3058]{Lauren M. Weiss}
\affiliation{Institute for Astronomy, University of Hawai`i, 2680 Woodlawn Drive, Honolulu, HI 96822, USA}

\author[0000-0002-2949-2163]{Edward W.\ Schwieterman}
\affiliation{Department of Earth and Planetary Sciences, University of California Riverside, 900 University Ave, Riverside, CA 92521, USA}
\affiliation{Blue Marble Space Institute of Science, Seattle, WA, 98115}

%%%%%%%%%%%%%%%%%%%%%%%%%%%%%%%%%%%%%%%%%%%%%%%%%%%%%%%%%%%%%%%%%%%%%%%%%%

\begin{abstract}
Despite the severe bias of the transit method of exoplanet discovery toward short orbital periods, a modest sample of transiting exoplanets with orbital periods greater than 100~days is known. Long-term radial velocity (RV) surveys are pivotal to confirming these signals and generating a set of planetary masses and densities for planets receiving moderate to low irradiation from their host stars. Here, we conduct RV observations of Kepler-1514 from the Keck I telescope using the High Resolution Echelle Spectrometer. From these data, we measure the mass of the statistically validated giant ({$1.108\pm0.023$~$R_{\rm J}$}) exoplanet Kepler-1514~b with a 218~day orbital period as {$5.28\pm0.22$~$M_{\rm J}$}. The bulk density of this cool ({$\sim$390~K}) giant planet is {$4.82^{+0.26}_{-0.25}$~g~cm$^{-3}$}, consistent with a core supported by electron degeneracy pressure. We also infer an orbital eccentricity of {$0.401^{+0.013}_{-0.014}$} from the RV and transit observations, which is consistent with planet-planet scattering and disk cavity migration models. The Kepler-1514 system contains an Earth-size, \textit{Kepler} Object of Interest on a 10.5~day orbit that we statistically validate against false positive scenarios, including those involving a neighboring star. The combination of the brightness ($V$=11.8) of the host star and the long period, low irradiation, and high density of Kepler-1514~b places this system among a rare group of known exoplanetary systems and one that is amenable to continued study. 
\end{abstract}

%%%%%%%%%%%%%%%%%%%%%%%%%%%%%%%%%%%%%%%%%%%%%%%%%%%%%%%%%%%%%%%%%%%%%%%%%%
 
\section{Introduction}\label{sec:intro}

The transit method is not conducive to the discovery of planets with orbital distances like those of the solar system planets. The probability of observing an exoplanet transit scales inversely with the star-planet separation due to geometry, from the random orientation of orbital inclinations, and sampling, from the limited baseline of continuous observations from transit surveys \citep{Beatty2008}. These factors have combined to largely exclude planets with orbital periods ($P$) greater than a hundred days from the list of known transiting exoplanets. 

The short-period bias of the transit method has a direct effect on the scientific return of observational investigations of exoplanets. The favorable geometry of a transit enables a suite of novel characterization techniques, most notably transmission spectroscopy \citep[e.g.,][]{Seager2000}. This technique has powered a thriving discipline of atmospheric characterization for short-period, close-in exoplanets \citep[e.g.,][]{Sing2016,Deming2017,Wakeford2017,Welbanks2019,Madhusudhan2019}. Similar observations, but of exoplanets on wider orbits with cooler temperatures would be equally as transformative and would enable new comparative studies between exoplanets and the solar system. Indeed, simulated observations of exoplanet analogs of the solar system giant planets have found an amenability to transmission spectroscopy \citep{Irwin2014,Dalba2015} as well as the novel technique of out-of-transit atmospheric characterization via refracted star light \citep{Sidis2010,Dalba2017b,Alp2018}.

Efforts to discover and maintain the ephemerides of long-period (roughly $P\gtrsim$100~days) transiting exoplanets have been underway for years. Some planets, like HD~80606~b, were first identified in radial velocity (RV) observations \citep{Naef2001} and were later found to have a transiting geometry \citep{Laughlin2009,Moutou2009}. However, this happy coincidence is expected to be quite rare \citep{Dalba2019a}. The vast majority of known long-period transiting exoplanets were identified through dedicated transit surveys. The constraints of ground-based observations have limited orbital periods of transiting exoplanets to less than roughly 25~days \citep[e.g.,][]{Brahm2016,Dittmann2017}. From space, where observational baselines are far less limited, a variety of exoplanets with orbital periods greater than approximately 100~days has been found.

Data from the primary \kepler\ mission \citep{Borucki2010,Thompson2018}---the longest continuous baseline transit survey conducted to date---have been meticulously searched for transits of long-period planets \citep{Wang2015,Morton2016,Uehara2016,Kawahara2019}. Related efforts have not only produced catalogs of objects with orbital periods between 100 and 1000 days, but have also revealed information about their underlying populations \citep{ForemanMackey2016b,Herman2019} and the likelihood of finding additional planets in their systems \citep{Dalba2016,Dalba2019c,Masuda2020}. A subset of \kepler's longest-period transiting planets are circumbinary \citep[e.g.,][]{Welsh2018,Socia2020} and are therefore amenable to a novel set of experiments and investigations.

Beyond Kepler, the repurposed {\it K2} mission \citep{Howell2014} also observed transits of a few planets and planet candidates with orbital periods on the order of hundreds of days despite its limited $\sim$75-day observational baseline between campaigns \citep{Osborn2016,Vanderburg2016b,Giles2018}. At even shorter observational baselines still, the ongoing Transiting Exoplanet Survey Satellite \citep[\tess;][]{Ricker2015} mission is contributing to the set of long-period exoplanets through single transit (or monotransit) events \citep{Cooke2018,Villanueva2019,Dalba2020a,Diaz2020,Eisner2020,Gill2020b,Lendl2020}. However, during TESS's primary mission, small patches of the sky (near the ecliptic poles) received near-continuous observations for almost a year. This strategy allows for the detection of two consecutive transits of an exoplanet with an orbital period on the order of 100~days. Moreover, TESS will observe many single-transit planet candidate host stars again during its extended mission and may detect additional transits that refine the ephemerides \citep[e.g.,][]{Cooke2020}.

Only a fraction of the exoplanets discovered in transit surveys are subject to follow-up mass measurement through RV monitoring. Stellar activity, rotational velocity, and the amplitude of RV variations induced by the planet relative to the precision of the facility are all factors that reduce the number of systems amenable to this characterization technique. The latter effect is crucial for long-period exoplanets as the RV semi-amplitude scales inversely with orbital period. There is also the issue that acquiring RV phase coverage for longer-period planets takes more time and requires longer-term stability of the facility. Yet, planetary confirmation through mass measurement is especially critical for giant planet candidates with $P\gtrsim$100~days that have been found to have a false-positive rate greater than 50\% in transit surveys \citep{Santerne2016}. However, since long-period orbits require long-duration follow-up campaigns, the number of long-period exoplanets with precise mass and radius is further limited \citep[e.g.,][]{Dubber2019}. 

Here, we add a new member to sample of exoplanets with $P>$100~days and precisely measured radii and masses: \planet\ (KOI~3681.01, KIC~2581316). \planet\ is a statistically validated, Jupiter-size planet \citep{Morton2016} that was found to have variations in the timing, depth, and duration of its transits \citep{Holczer2016}. The \host\ system also contains a \kepler\ Object of Interest (KOI) planet candidate, \koi, with a shallower transit and a 10.5~d orbital period, which we validate as Kepler-1514~c. \host\ therefore joins the list of systems with interior Earth-sized or super-Earth-sized exoplanets with exterior giant planet companions \citep[e.g.,][]{Zhu2018,Bryan2019}. The host star itself has a $V$-band magnitude of 11.8, which is brighter than 96\% of other stars with planets on long-period ($P>$100~days) orbits discovered by \kepler. 

The rest of this paper is organized as follows. In Section \ref{sec:obs}, we describe the photometry of the \host\ system from the primary \kepler\ mission and our spectroscopic follow-up observations from the Keck I telescope. In Section \ref{sec:model}, we conduct a global modeling of the photometric and spectroscopic data to infer the various stellar, planetary, and orbital properties of the objects in the \host\ system. Also, we tailor our approach to investigate how the observed rotational variability of \host\ affects the inferred transit properties of \planet. In Section \ref{sec:results}, we confirm the planetary nature of \planet\ by measuring its mass and we statistically validate \koi. In Section \ref{sec:disc}, we discuss the properties \planet\ and its host star relative to the sample of other weakly-irradiated, cool giant exoplanets. Finally, in Section \ref{sec:conc}, we summarize our findings.

%%%%%%%%%%%%%%%%%%%%%%%%%%%%%%%%%%%%%%%%%%%%%%%%%%%%%%%%%%%%%%%%%%%%%%%%%%

\section{Observations}\label{sec:obs}

We employ photometric, spectroscopic, and imaging observations in this analysis of the \host\ system. In the following sections, we describe how each of these data sets was collected and processed.

%%%%%%%%%%%%%%%%%%%%%%%%%%%%%%%%%%%%%%%%%%%%%%%%%%%%%%%%%%%%%%%%%%%%%%%%%%

\subsection{Photometric Data from Kepler}\label{sec:kepler}

The \kepler\ spacecraft observed \host\ in 18 quarters of its primary mission. These observations captured seven transits of the outer planet \planet\ and over 100 transits of the inner planet candidate \koi. We accessed the simple aperture photometry (SAP) and pre-search data conditioning (PDC) light curves \citep{Jenkins2010a,Smith2012,Stumpe2012} from \kepler\ through the Milkuski Archive for Space Telescopes (MAST). Both types of photometry contain significant brightness variations. The SAP light curves contain systematic variations induced by spacecraft motion as well as stellar variability while the PDC light curves contain variations introduced by the detrending. In either case, special consideration is required to model the transit events. We proceed with the SAP data products to ensure that the PDC systematics correction does not distort the deep, long-duration transits of \planet. The crowding metric for each quarter is $\sim$1 suggesting that the \kepler\ photometric apertures and resulting radius measurements are not contaminated by background sources (also see Section \ref{sec:imaging}). We also verify that the apertures are not contaminated by so-called ``phantom stars,'' which are non-existent sources often resulting from errors in all-sky photometric catalogs \citep{Dalba2017a}. 

\begin{figure}
    \centering
    \includegraphics[width=0.9\columnwidth]{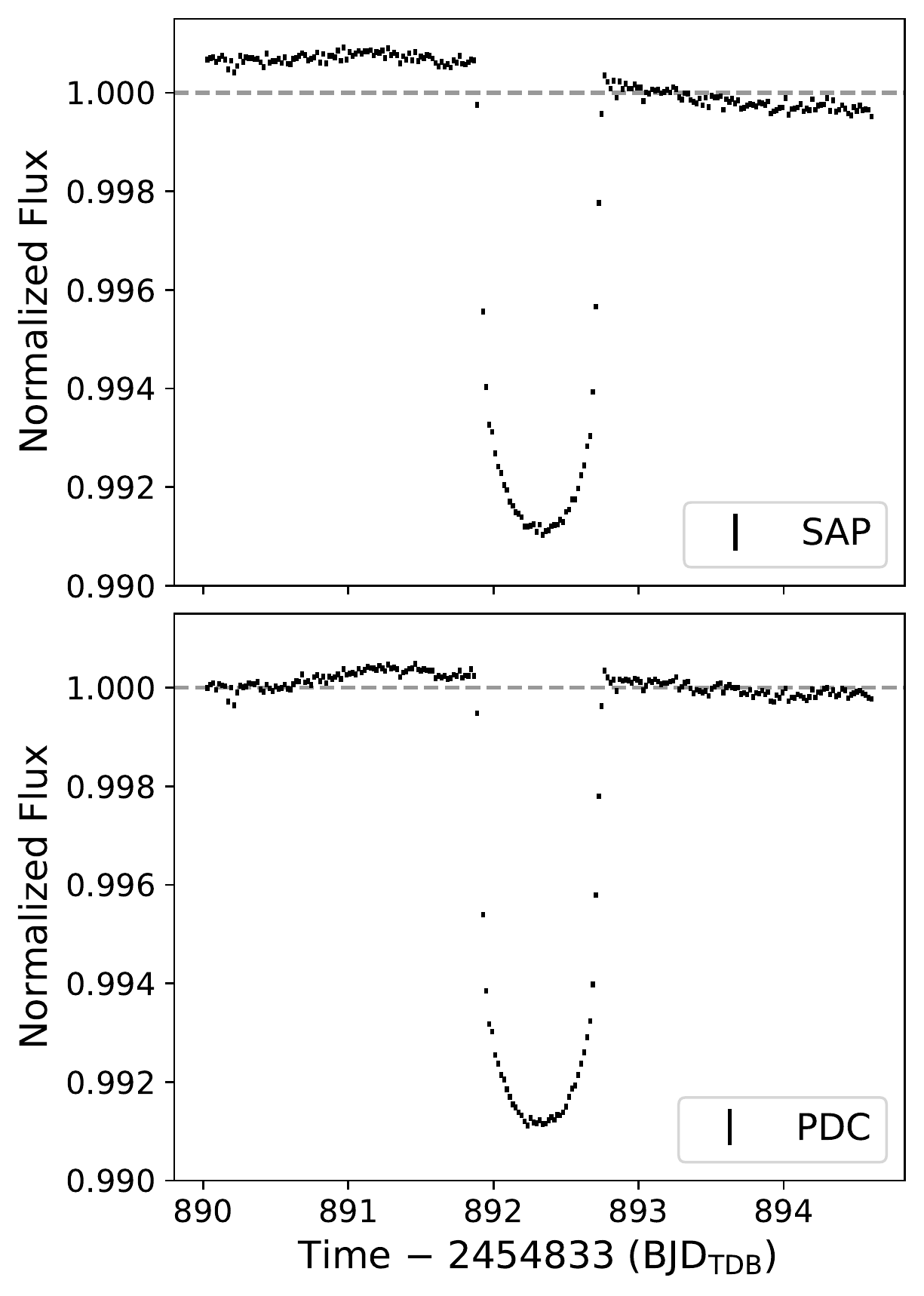}
    \caption{Median-normalized, transit light curve of \planet\ from Quarter 9 using \kepler\ SAP (top) and PDC (bottom) data products. We explore whether the variability that is present in these light curves could account for the TTVs, T$\delta$Vs, and TDVs measured by \citet{Holczer2016} in our modeling of this system.}
    \label{fig:pdc-sap}
\end{figure}

In Figure \ref{fig:pdc-sap}, we show the Quarter 9 transit of \planet\ to illustrate the typical level of variability present in the SAP and PDC light curves. A previous analysis of the \kepler\ PDC photometry of \host\ measured variations in transit timing (TTV), depth (T$\delta$V), and duration (TDV) for \planet, although the statistical significance of these measurements were low \citep{Holczer2016}. Stellar variability, including brightness variations caused by spots, can cause transit ephemeris variations \citep[e.g.,][]{Alonso2008,Oshagh2013}. \citet{Holczer2016} employed a photometric detrending algorithm to prevent the false detection of TTVs due to stellar variability, but their efforts were spread across a wide catalog of stars and transiting planets. The low statistical significance of the purported transit variations combined with the variability present in the \kepler\ light curves of \host\ warrant the focused detrending procedures that we employ in Section \ref{sec:model}. 

%%%%%%%%%%%%%%%%%%%%%%%%%%%%%%%%%%%%%%%%%%%%%%%%%%%%%%%%%%%%%%%%%%%%%%%%%%

\subsection{Spectroscopic Data from HIRES}\label{sec:hires}

We acquired 12 high resolution spectra of \host\ with the High Resolution Echelle Spectrometer \citep[HIRES;][]{Vogt1994} on the Keck I telescope. One spectrum was acquired with a high signal-to-noise ratio (S/N) of $\sim$190 without a heated iodine in the light path. This spectrum is used for a spectroscopic analysis of \host\ and is vetted for a second set of spectral lines following the methods of \citet{Kolbl2015}. We rule out additional spectral lines brighter than 1\% of the primary's and at velocity separations greater than 10~km~s$^{-1}$. This high S/N spectrum also served as a spectral template in the standard forward modeling procedures employed by the California Planet Search \citep[e.g.,][]{Howard2010,Howard2016}, thereby removing the need to synthesize a spectral template \citep{Fulton2015b} or match \host\ to another star in the HIRES template library \citep{Dalba2020b}. The RVs are listed in Table \ref{tab:rvs}. Since the HIRES spectra include the Ca II H and K spectral lines, each value of RVs is accompanied by a correspond $S_{\rm HK}$ activity indicator \citep{Isaacson2010}.

\begin{deluxetable}{ccc}
\tablecaption{RV Measurements of \host. \label{tab:rvs}}
\tablehead{
  \colhead{BJD$_{\rm TDB}$} & 
  \colhead{RV (m s$^{-1}$)} &
  \colhead{$S_{\rm HK}$}}
\startdata
2458346.85153 & $40.6\pm4.3$   & $0.139\pm0.001$ \\
2458361.02310 & $12.5\pm4.4$   & $0.141\pm0.001$ \\
2458390.72137 & $-56.7\pm3.9$  & $0.140\pm0.001$ \\
2458396.76976 & $-68.6\pm5.0$  & $0.140\pm0.001$ \\
2458560.14495 & $39.2\pm4.2$   & $0.135\pm0.001$ \\
2458622.94024 & $-85.3\pm3.8$  & $0.145\pm0.001$ \\
2458650.97962 & $-113.1\pm4.0$ & $0.146\pm0.001$ \\
2458663.07909 & $-97.8\pm4.2$  & $0.142\pm0.001$ \\
2458737.82511 & $158.4\pm4.3$  & $0.131\pm0.001$ \\
2458787.84946 & $25.3\pm3.8$   & $0.135\pm0.001$ \\
2458906.15457 & $141.3\pm3.8$  & $0.128\pm0.001$ \\
\enddata
\end{deluxetable}

%%%%%%%%%%%%%%%%%%%%%%%%%%%%%%%%%%%%%%%%%%%%%%%%%%%%%%%%%%%%%%%%%%%%%%%%%%

\subsection{Archival Imaging Data from NIRC2}\label{sec:imaging}

\host\ was observed at high angular resolution by \citet{Kraus2016} on 2014 August 12 using the NIRC2 adaptive optics imager at Keck Observatory \citep{Wizinowich2000}. The observation used adaptive optics imaging, coronagraphy, and non-redundant aperture mask interferometry to reveal a neighbor located $\rho = $0$\farcs$272 away from the apparent planet-hosting star with an apparent contrast of $\Delta K^{\prime} = 6.06$~mag, while also achieving deep and close limits for any additional neighbors that might account for the transit signals. This system was also observed with speckle imaging at visible wavelengths at the Wisconsin-Indiana-Yale-NOAO (WIYN) telescope using the DSSI speckle camera \citep{Furlan2017a}. The neighbor was not detected, but at 0$\farcs$27 projected separation, the speckle observations yielded relative contrast limits of $\Delta m_{692} = 3.05$~mag and $\Delta m_{880} = 2.50$~mag.

\host\ was also observed with Keck-II/NIRC2 on 2013 July 7 (as reported by \citealt{Furlan2017a}) and on 2015 July 26 (PI Dupuy). The proper motion of \host\ is $\mu = 10$~mas~yr$^{-1}$, while NIRC2 astrometry of close binary pairs can be measured with a precision of $\la$1--2~mas \citep[e.g.,][]{Dupuy2016}, so the two year baseline offers the opportunity to distinguish whether the neighbor is a comoving low-mass companion, or a chance alignment with a background star. We therefore have analyzed the images from all three epochs using the same methods described in \citet{Kraus2016}. To briefly recap, our pipeline fits each image of the close pair with a double point spread function (PSF) model based in the best-fitting single star PSF selected from all those observed nearby in time, and then the relative astrometry is corrected for the known optical distortion of NIRC2 \citep{Yelda2010}. 

\begin{figure}
    \centering
    \includegraphics[width=\columnwidth]{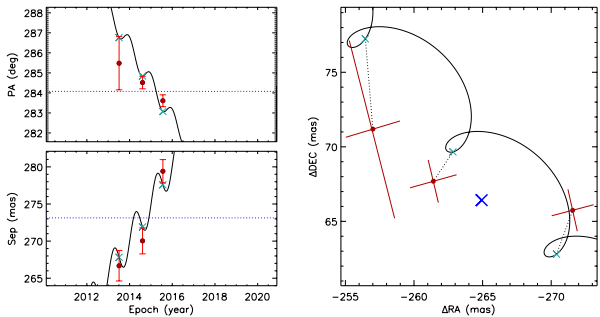}
    \caption{Relative motion of the close neighbor to \host, as measured from multi-epoch astrometry using adaptive optics imaging. The left panels show the separation and position angle between \host\ and its neighbor as a function of time, while the right panel shows the relative motion of the neighbor in the plane of the sky. The expected trajectory of a non-moving background star is shown with the solid curve, while the expected relative position of a comoving binary companion is shown with dotted lines in the left panels and a blue X in the right panel. We conclude that the faint neighbor is not bound to \host, and is instead a chance alignment with an unrelated field interloper.}
    \label{fig:imaging}
\end{figure}

In Table~\ref{tab:imaging}, we summarize the relative astrometry and photometry that we measured at each epoch, computing a simple mean of the fit results from the individual images. In Figure~\ref{fig:imaging}, we plot the corresponding relative motion over time, also showing the trajectories expected for a completely comoving neighbor or a completely non-moving background star. We find that the background star solution is consistent with the observations ($\chi^2 = 8.1$ on 4 degrees of freedom; $P = 0.09$), whereas the comoving solution is inconsistent with the observations ($\chi^2 = 34.6$ on 4 degrees of freedom; $P = 5 \times 10^{-7}$). The escape velocity of a bound companion at a projected separation of $\rho = 0\farcs272$ or $\rho = 110$~au would only be $\Delta v_{esc} \sim 3$~km~s$^{-1}$ or $\Delta \mu_{esc} \sim 1.5$~mas~yr$^{-1}$, much lower than the measured relative motion. We therefore conclude that the relative motion can not be orbital motion and the neighbor is a field star seen in chance alignment, not a bound binary companion.

\begin{deluxetable*}{ccccccc}
\tabletypesize{\footnotesize}
\tablewidth{0pt}
\tablecaption{Summary of \host\ Neighbor Detections from NIRC2 PSF Fitting \label{tab:imaging}}
\tablehead{
\colhead{Epoch} & \colhead{Filter} & \colhead{$N_{obs}$} &  \colhead{$\rho$} & \colhead{$PA$} & \colhead{$\Delta m$} &  \colhead{PI} \\
\colhead{(MJD)} & \colhead{} & \colhead{} & \colhead{(mas)} & \colhead{(deg)} & \colhead{(mag)} 
}
\startdata
56480.53 &  Brg &  11 &  266.68 $\pm$ 2.05 &  285.482 $\pm$ 1.328 &  6.130 $\pm$0.164 & Weaver \\ % D N2.20130707.45389
56881.51 &   Kp &   2 &  270.03 $\pm$ 1.75 &  284.517 $\pm$ 0.313 &  6.062 $\pm$0.033 & Kraus \\ % D N2.20140812.44077
57229.56 &   Kp &   6 &  279.41 $\pm$ 1.57 &  283.609 $\pm$ 0.296 &  6.240 $\pm$0.060 & Dupuy \\ % D N2.20150726.47994
\enddata
\end{deluxetable*}

Distant background stars are likely to be relatively blue early-type dwarfs, so the contrast in the \kepler\ bandpass is likely to be similar to that in the near-infrared ($\Delta K^{\prime} = 6$~mag). Under this assumption, the transit depth is only diluted by 0.4\%, leading to a planet radius change of 0.2\%, well within the measured uncertainty. Therefore, we hereafter neglect any flux contribution that this neighbor made in the transit fits, and we show in Section \ref{sec:koi} that the signal from \koi\ cannot originate from this faint field interloper.

%%%%%%%%%%%%%%%%%%%%%%%%%%%%%%%%%%%%%%%%%%%%%%%%%%%%%%%%%%%%%%%%%%%%%%%%%%

\section{Modeling Stellar and Planetary Parameters}\label{sec:model}

We conducted joint modeling of the stellar, transit, and RV data of \host\ to infer various stellar, planetary, and systemic parameters using the \textsf{EXOFASTv2} modeling suite \citep{Eastman2013,Eastman2017,Eastman2019}. Since the photometric variability tied to the rotation of \host\ can affect the derived transit parameters, we first applied special detrending to remove this rotational modulation. Then, we conducted an initial \textsf{EXOFASTv2} fit to assess the impact of this detrending on the variations in transit parameters previously measured for \planet. Finally, we ran a comprehensive \textsf{EXOFASTv2} fit that models the \planet\ and the \koi\ from which we derive the final system parameters.

%%%%%%%%%%%%%%%%%%%%%%%%%%%%%%%%%%%%%%%%%%%%%%%%%%%%%%%%%%%%%%%%%%%%%%%%%%

\subsection{Removal of Out-of-transit Photometric Variability}\label{sec:spline}

The SAP light curves contain long-term variations due to stellar activity and instrumental drifts. These are dominated by differential velocity aberration (DVA), which is the change in the local pixel scale and distortion of the scene caused by spacecraft motion \citep[e.g.,][]{Kinemuchi2012}. DVA yields a linear or quadratic slope over the duration of a \kepler\ quarter that is negligible on the 21~hr timescale of transit. We modeled these variations with a basis spline which we fit simultaneously with the shape of the two transit signals for \planet\ and \koi. Our strategy is similar to that of \citet{Vanderburg2016a}, except that we do not also model spacecraft systematic noise in our well-behaved \kepler\ data\footnote{\url{https://github.com/avanderburg/keplerspline}.}. In brief, we started by clipping anomalous data taken during the following time intervals (given in BKJD, or BJD $-$ 2454833): $247 <t< 260$, $1160.5 < t < 1162$, and $1289 < t < 1296$. We identified all gaps in the light curve longer than 0.3 days and introduced discontinuities in our spline at these points. We modeled the two transit signals with analytic \citet{Mandel2002} curves and minimized $\chi^2$ with a Levenberg-Marquardt algorithm \citep{Markwardt2009}. At each step of the minimization, we calculated the transit models, subtracted them from the light curve, and then fit the basis spline to this residual curve. We then minimized the deviations of (data $-$ transit model $-$ spline). After the optimization concluded, we calculated a final spline from the residuals to the best-fit transit model and subtracted it from the light curve to remove the long-term variability.

%%%%%%%%%%%%%%%%%%%%%%%%%%%%%%%%%%%%%%%%%%%%%%%%%%%%%%%%%%%%%%%%%%%%%%%%%%

\subsection{Preliminary EXOFASTv2 Modeling}\label{sec:prelim}

After detrending the light curves of \host, we completed a preliminary model fit to the transit and RV data using \textsf{EXOFASTv2}. The purpose of this fit was to determine if the detrending affected the TTVs and T$\delta$Vs measured previously by \citet{Holczer2016}, so we allowed extra parameters describing the timing and depth of each transit. We did not investigate TDVs as the values measured by \citet{Holczer2016} are fully consistent with no variation in transit duration. We only included transits of \planet\ in the fit. The fit converged according to the default \textsf{EXOFASTv2} statistics for each parameter: the number of independent draws of the underlying posterior probability distribution \citep[$T_z>1000$,][]{Ford2006b} and the well known Gelman--Rubin statistic \citep[GR$<1.01$,][]{Gelman1992}.

\begin{figure}
    \centering
    \includegraphics[width=\columnwidth]{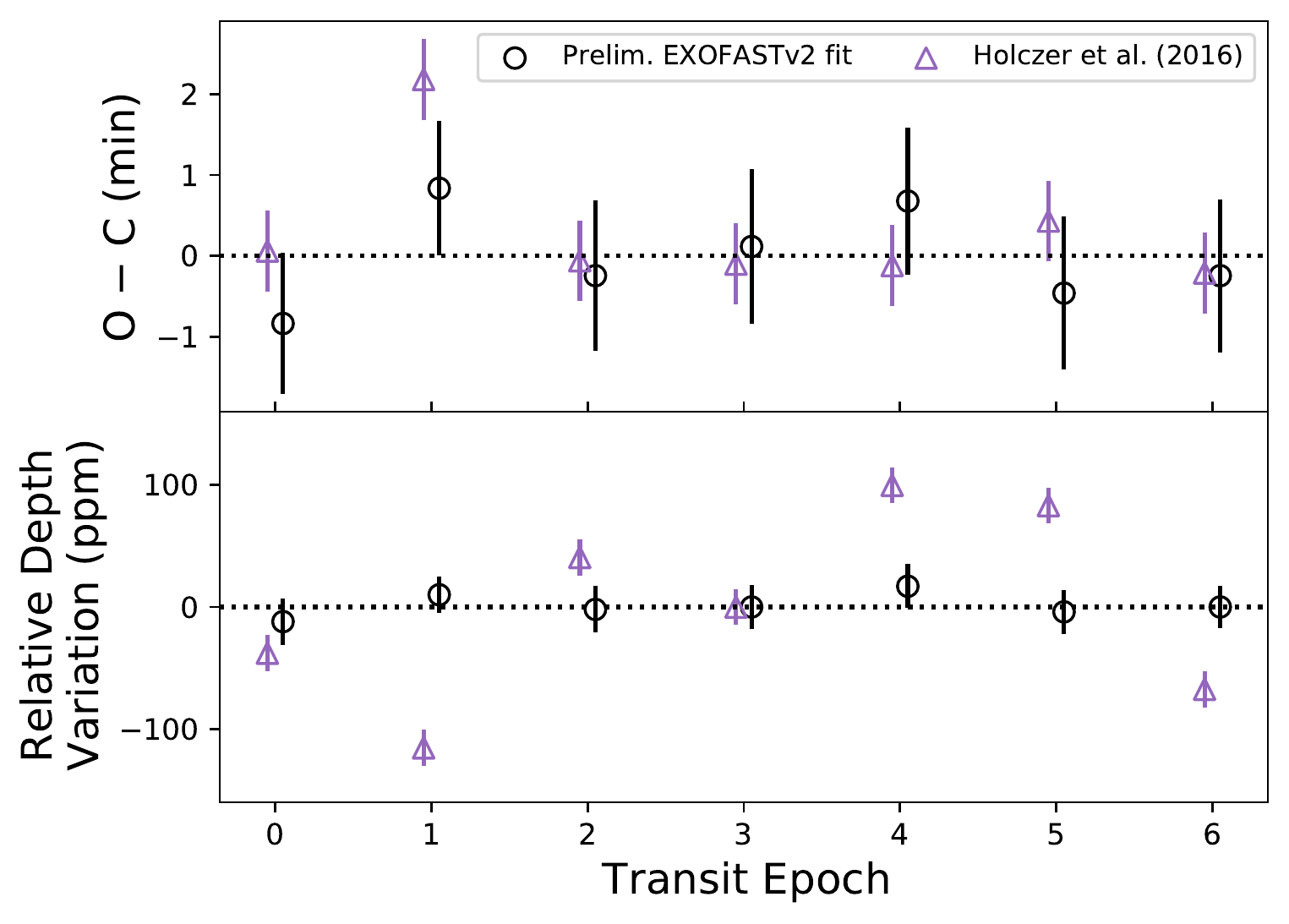}
    \caption{Observed minus calculated (O $-$ C) timing of the transits (top) and transit depth variations fit relative to the first transit and then median-subtracted (bottom) of \planet\ from the preliminary \textsf{EXOFASTv2} fit (Section \ref{sec:prelim}). The data sets have been offset horizontally for clarity. In both panels, corresponding values from \citet{Holczer2016} are shown. When detrending the light curves with a spline, we find that the transit depth variations become insignificant.}
    \label{fig:ttvs1}
\end{figure}

We show the values of TTVs and T$\delta$Vs inferred from this preliminary modeling along with those values from \citet{Holczer2016} in Figure \ref{fig:ttvs1}. The TTVs are presented as the difference between the observed ephemeris and the calculated (linear) ephemeris (i.e., O$-$C). The T$\delta$Vs were fit relative to the first transit but are shown as median-subtracted values in Figure \ref{fig:ttvs1}. The TTVs we measure are consistent with, although slightly less precise than, those reported by \citet{Holczer2016}. We quantify their significance as the reduced $\chi^2$ statistic when compared to a linear ephemeris (i.e., a flat line at O~$-$~C = 0), which equals 0.5. Although weak, we cannot claim that these TTVs are negligible nor can we distinguish between photometric variability or dynamical interaction as their cause. Consequently, we decide to include TTVs in the comprehensive modeling the of the \host\ system data.

On the other hand, we do not detect T$\delta$Vs in the \planet\ transits, a result that is inconsistent with \citet{Holczer2016}. This discrepancy suggests photometric detrending as the probable cause of the purported T$\delta$Vs. On this basis, we do not include T$\delta$Vs in the modeling of the \host\ system hereafter.

%%%%%%%%%%%%%%%%%%%%%%%%%%%%%%%%%%%%%%%%%%%%%%%%%%%%%%%%%%%%%%%%%%%%%%%%%%

\subsection{Final, Comprehensive EXOFASTv2 Modeling}\label{sec:final}

\begin{figure*}
    \centering
    \includegraphics[width=0.9\textwidth]{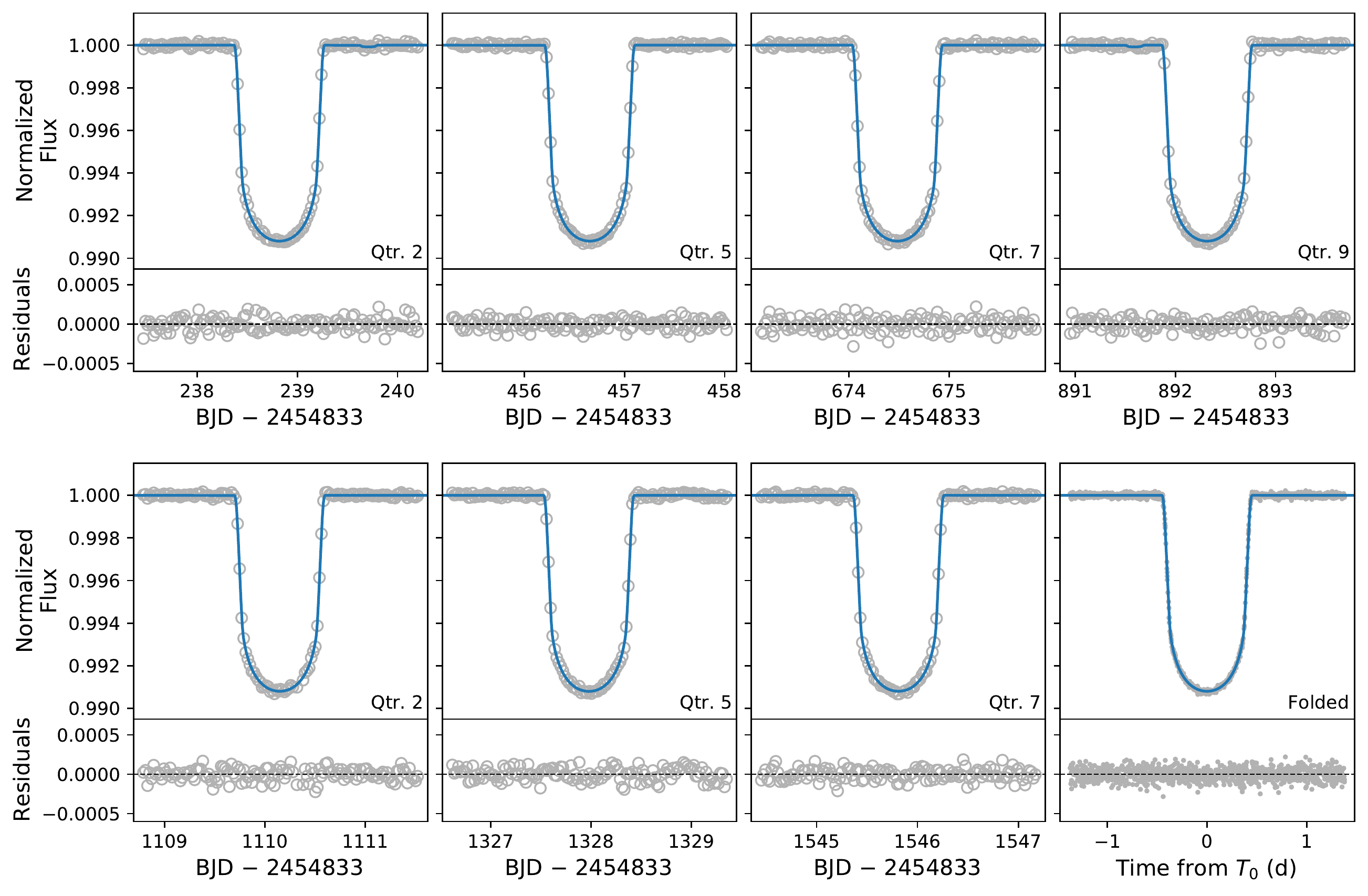}
    \caption{All long cadence transits of \planet, labeled by \kepler\ Quarter, and then folded on the best-fit ephemeris in the bottom-right panel. The blue lines are the best-fit model, which includes TTVs but not T$\delta$Vs.}
    \label{fig:b_transits}
\end{figure*}

\begin{figure}
    \centering
    \includegraphics[width=\columnwidth]{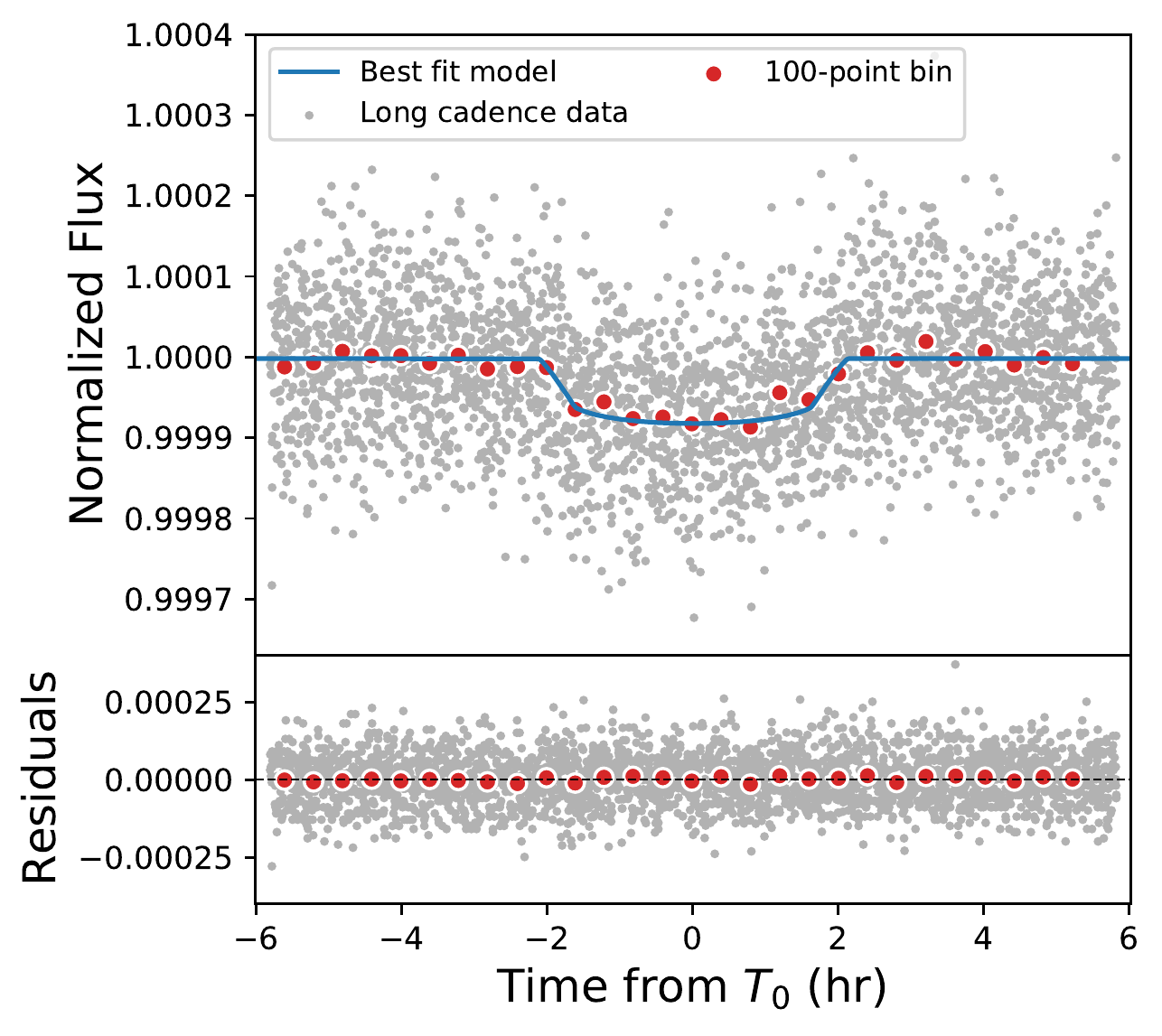}
    \caption{\kepler\ long cadence transits of \koi\ folded on the best-fit ephemeris, which does not include TTVs. The binned data clearly identify the shallow transit of the exoplanet candidate.}
    \label{fig:koi_transits}
\end{figure}

\begin{figure}
    \centering
    \includegraphics[width=\columnwidth]{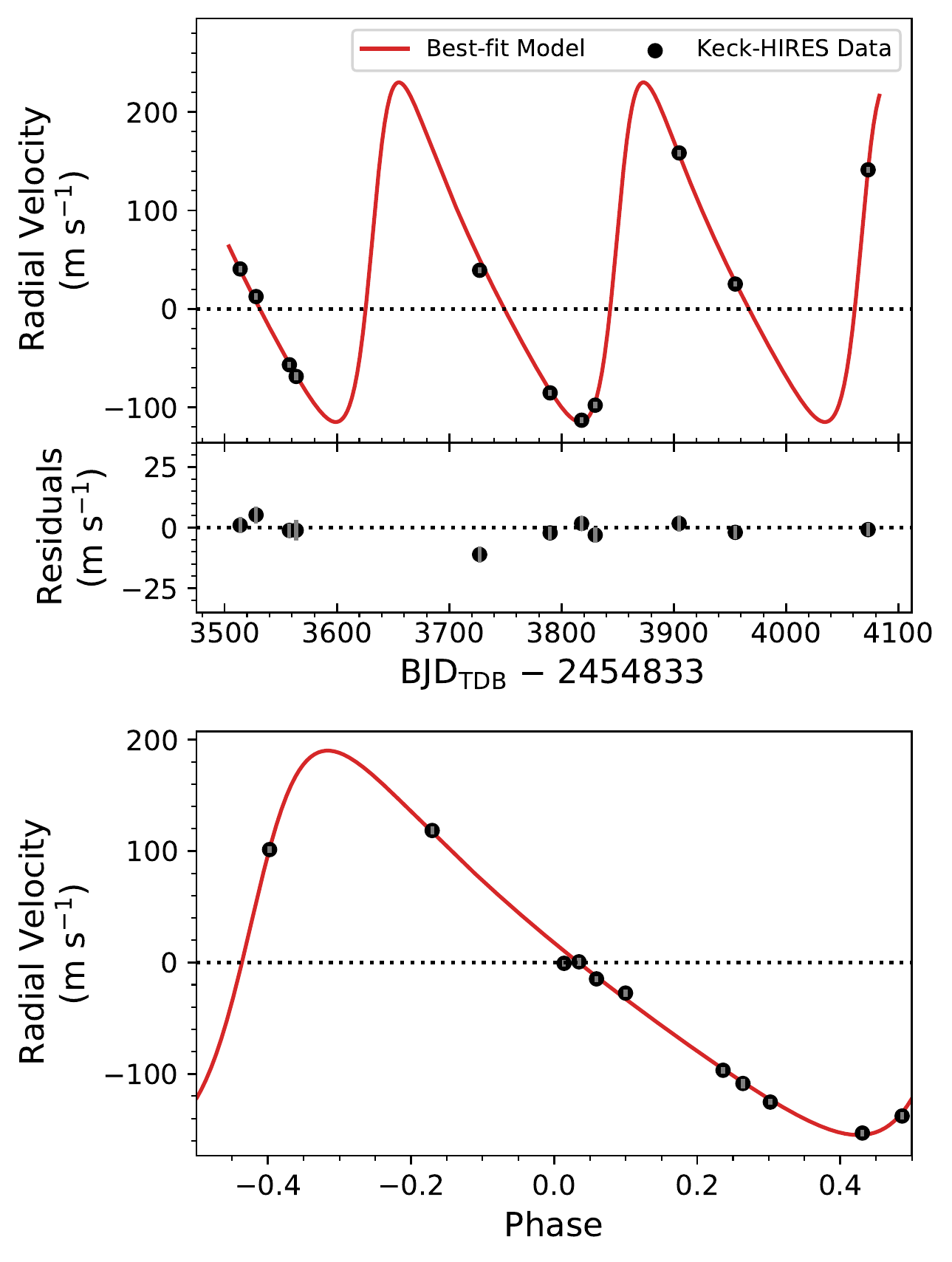}
    \caption{RV measurements of \host\ from Keck-HIRES. The top panel is the time series data and the bottom panel shows the data phase folded on the best-fit ephemeris using the time of conjunction ($T_C$) as the reference point. Error bars are small but are shown in gray in each panel.}
    \label{fig:rv}
\end{figure}

For the final global analysis presented in Tables \ref{tab:stellar} and \ref{tab:planets}, we conduct the \textsf{EXOFASTv2} fit in the following fashion. We jointly fit the available detrended \kepler\ light curve for both planets, but we only fit the Keck-HIRES RVs and allow for TTVs for \planet. We exclude fitting the RVs for \koi\ since the measured size from our fit ({1.15~$R_{\earth}$}) suggests a planet mass on the order of $\sim$1~$M_{\earth}$. A 1~$M_{\earth}$ planet on a circular orbit which would produce an RV semi-amplitude of $\sim$26~cm~s$^{-1}$, which is below the internal precision of the Keck-HIRES measurements and may not be detectable with any amount of data. Within the fit, the host star parameters were determined using the spectral energy distribution (SED) from broadband photometry and the MESA Isochrones and Stellar Tracks (MIST) stellar evolution models \citep{Paxton2011, Paxton2013, Paxton2015, Choi2016, Dotter2016}. We place a Gaussian prior of 2.5705$\pm$0.0418~mas on parallax based on measurements from Gaia \citep{Gaia2018}, which we correct for the offset reported by \citet{Stassun2018a}. We also place a Gaussian prior on the stellar metallicity ($[$Fe/H$]$=0.05$\pm$0.09) based on spectroscopic analysis of the high S/N template spectrum following \citet{Yee2017}. Lastly, we employ an upper limit on the line of sight extinction (A$_V<$0.5115) from the \citet{Schlegel1998} galactic dust maps. We allow the fit to proceed until convergence as quantified by at least 1000 independent draws from the posterior probability distribution of each fitted parameter \citep{Ford2006b} and by a Gelman--Rubin statistic of less than or equal to 1.01 for each fitted parameter \citep{Gelman1992}. The stellar and planetary parameters inferred from the comprehensive \textsf{EXOFASTv2} modeling are listed Table \ref{tab:stellar} and \ref{tab:planets}, respectively. The final transit and RV data sets along with the best-fit models for the \host\ system are presented in Figures \ref{fig:b_transits}, \ref{fig:koi_transits}, and \ref{fig:rv}. 

%Stellar param table
\begin{deluxetable}{lcc}
\tablecaption{Median values and 68\% confidence intervals for Kepler-1514 stellar parameters \label{tab:stellar}}
\tablehead{\colhead{~~~Parameter} & \colhead{Units} & \colhead{Values}}
\startdata
\multicolumn{2}{l}{Informative Priors:}&\smallskip\\
~~~~$[{\rm Fe/H}]$\dotfill &Metallicity (dex)\dotfill &$\mathcal{N}(0.05,0.09)$\\
~~~~$\varpi$\dotfill &Parallax (mas)\dotfill &$\mathcal{N}(2.5705,0.0418)$\\
~~~~$A_V$\dotfill &V-band extinction (mag)\dotfill &$\mathcal{U}(0,0.5115)$\\
\smallskip\\\multicolumn{2}{l}{Stellar Parameters:}&\smallskip\\
~~~~$M_*$\dotfill &Mass (\msun)\dotfill &$1.196^{+0.065}_{-0.063}$\\
~~~~$R_*$\dotfill &Radius (\rsun)\dotfill &$1.289^{+0.027}_{-0.026}$\\
~~~~$L_*$\dotfill &Luminosity (\lsun)\dotfill &$2.13^{+0.16}_{-0.12}$\\
~~~~$F_{Bol}$\dotfill &Bolometric Flux (cgs)\dotfill &$4.49^{+0.31}_{-0.20}\times10^{-10}$\\
~~~~$\rho_*$\dotfill &Density (g~cm$^{-3}$)\dotfill &$0.787^{+0.041}_{-0.040}$\\
~~~~$\log{g}$\dotfill &Surface gravity (cgs)\dotfill &$4.295\pm0.019$\\
~~~~$T_{\rm eff}$\dotfill &Effective Temperature (K)\dotfill &$6145^{+99}_{-80}$\\
~~~~$[{\rm Fe/H}]$\dotfill &Metallicity (dex)\dotfill &$0.119^{+0.080}_{-0.075}$\\
~~~~$[{\rm Fe/H}]_{0}$\dotfill &Initial Metallicity$^{a}$ \dotfill &$0.163^{+0.066}_{-0.064}$\\
~~~~$Age$\dotfill &Age (Gyr)\dotfill &$2.9^{+1.6}_{-1.3}$\\
~~~~$EEP$\dotfill &Equal Evolutionary Phase$^{b}$ \dotfill &$361^{+34}_{-24}$\\
~~~~$A_V$\dotfill &V-band extinction (mag)\dotfill &$0.076^{+0.077}_{-0.053}$\\
~~~~$\sigma_{SED}$\dotfill &SED photometry error scaling \dotfill &$0.70^{+0.25}_{-0.16}$\\
~~~~$\varpi$\dotfill &Parallax (mas)\dotfill &$2.568\pm0.040$\\
~~~~$d$\dotfill &Distance (pc)\dotfill &$389.3^{+6.1}_{-5.9}$\\
\smallskip\\\multicolumn{2}{l}{Wavelength Parameters:}&Kepler\smallskip\\
~~~~$u_{1}$\dotfill &linear limb-darkening coeff \dotfill &$0.3474^{+0.0076}_{-0.0077}$\\
~~~~$u_{2}$\dotfill &quadratic limb-darkening coeff \dotfill &$0.248\pm0.016$\\
\enddata
\tablenotetext{}{See Table 3 in \citet{Eastman2019} for a detailed description of all parameters and all default (non-informative) priors beyond those specified here.}
\tablenotetext{a}{Initial metallicity is that of the star when it formed.}
\tablenotetext{b}{Corresponds to static points in a star's evolutionary history. See Section~2 in \citet{Dotter2016}.}
\end{deluxetable}

%Planet param table
\begin{deluxetable*}{lccc}
\tablecaption{Median values and 68\% confidence interval for the planets in the Kepler-1514 System\label{tab:planets}}
\tablehead{\colhead{~~~Parameter} & \colhead{Units} & \multicolumn{2}{c}{Values}}
\startdata
\multicolumn{2}{l}{Planetary Parameters:}&b&c\smallskip\\
~~~~$P$\dotfill &Period (days)\dotfill &$217.83184\pm0.00012$&$10.514181\pm0.000039$\\
~~~~$R_P$\dotfill &Radius (\rj)\dotfill &$1.108\pm0.023$&$0.1049^{+0.0051}_{-0.0039}$\\
~~~~$M_P$\dotfill &Mass (\mj)\dotfill &$5.28\pm0.22$&\nodata\\
~~~~$T_C$\dotfill &Time of conjunction$^{a}$ (\bjdtdb)\dotfill &$2455071.81411\pm0.00046$&$2454957.0546^{+0.0034}_{-0.0036}$\\
~~~~$a$\dotfill &Semi-major axis (AU)\dotfill &$0.753^{+0.013}_{-0.014}$&$0.0997\pm0.0018$\\
~~~~$i$\dotfill &Inclination (Degrees)\dotfill &$89.944^{+0.013}_{-0.010}$&$87.98^{+1.2}_{-0.40}$\\
~~~~$e$\dotfill &Eccentricity$^{b}$ \dotfill &$0.401^{+0.013}_{-0.014}$&$0.32^{+0.35}_{-0.19}$\\
~~~~$\omega_*$\dotfill &Argument of Periastron (Degrees)\dotfill &$-75.28^{+0.75}_{-0.71}$&$0^{+120}_{-160}$\\
~~~~$T_{eq}$\dotfill &Equilibrium temperature$^{c}$ (K)\dotfill &$387.9^{+6.0}_{-5.0}$&$1066^{+16}_{-14}$\\
~~~~$K$\dotfill &RV semi-amplitude (m~s$^{-1}$)\dotfill &$172.5\pm3.9$&\nodata\\
~~~~$R_P/R_*$\dotfill &Radius of planet in stellar radii \dotfill &$0.08835^{+0.00014}_{-0.00015}$&$0.00836^{+0.00037}_{-0.00026}$\\
~~~~$a/R_*$\dotfill &Semi-major axis in stellar radii \dotfill &$125.6\pm2.2$&$16.63\pm0.29$\\
~~~~$\delta$\dotfill &Transit depth (fraction)\dotfill &$0.007805^{+0.000025}_{-0.000027}$&$0.0000699^{+0.0000063}_{-0.0000043}$\\
~~~~$\tau$\dotfill &Ingress/egress transit duration (days)\dotfill &$0.07409^{+0.00093}_{-0.00097}$&$0.00166^{+0.00091}_{-0.00036}$\\
~~~~$T_{14}$\dotfill &Total transit duration (days)\dotfill &$0.88862^{+0.00077}_{-0.00078}$&$0.1567^{+0.0035}_{-0.0034}$\\
~~~~$b$\dotfill &Transit Impact parameter \dotfill &$0.169^{+0.030}_{-0.039}$&$0.47^{+0.22}_{-0.31}$\\
~~~~$b_S$\dotfill &Eclipse impact parameter \dotfill &$0.074^{+0.013}_{-0.017}$&$0.44^{+0.18}_{-0.28}$\\
~~~~$\tau_S$\dotfill &Ingress/egress eclipse duration (days)\dotfill &$0.0323\pm0.0011$&$0.00195^{+0.00035}_{-0.00083}$\\
~~~~$T_{S,14}$\dotfill &Total eclipse duration (days)\dotfill &$0.395\pm0.013$&$0.160^{+0.058}_{-0.042}$\\
~~~~$\rho_P$\dotfill &Density (g~cm$^{-3}$)\dotfill &$4.82^{+0.26}_{-0.25}$&\nodata\\
~~~~$log\;g_P$\dotfill &Surface gravity \dotfill &$4.028\pm0.017$&\nodata\\
~~~~$\fave$\dotfill &Incident Flux (\fluxcgs)\dotfill &$0.00440^{+0.00029}_{-0.00024}$&$0.263^{+0.029}_{-0.066}$\\
~~~~$T_P$\dotfill &Time of Periastron (\bjdtdb)\dotfill &$2454981.75^{+0.73}_{-0.74}$&$2454956.9\pm1.3$\\
~~~~$T_S$\dotfill &Time of eclipse (\bjdtdb)\dotfill &$2455196.06^{+0.61}_{-0.63}$&$2454951.8^{+2.7}_{-2.9}$\\
~~~~$T_A$\dotfill &Time of Ascending Node (\bjdtdb)\dotfill &$2455002.91^{+0.82}_{-0.78}$&$2454954.8^{+1.3}_{-1.5}$\\
~~~~$T_D$\dotfill &Time of Descending Node (\bjdtdb)\dotfill &$2455164.5\pm1.2$&$2454959.3^{+1.5}_{-1.3}$\\
~~~~$e\cos{\omega_*}$\dotfill & \dotfill &$0.1021^{+0.0040}_{-0.0041}$&$0.00\pm0.42$\\
~~~~$e\sin{\omega_*}$\dotfill & \dotfill &$-0.388\pm0.014$&$0.02^{+0.17}_{-0.27}$\\
~~~~$P_S$\dotfill &A priori non-grazing eclipse prob \dotfill &$0.01201\pm0.00011$&$0.063^{+0.065}_{-0.011}$\\
~~~~$P_{S,G}$\dotfill &A priori eclipse prob \dotfill &$0.01434\pm0.00013$&$0.064^{+0.066}_{-0.011}$\\
\smallskip\\\multicolumn{2}{l}{Telescope Parameters:}&Keck~I\smallskip\\
~~~~$\gamma_{\rm rel}$\dotfill &Relative RV Offset (m~s$^{-1}$)\dotfill &$38.9\pm2.1$\\
~~~~$\sigma_J$\dotfill &RV Jitter (m~s$^{-1}$)\dotfill &$4.2^{+3.2}_{-2.8}$\\
~~~~$\sigma_J^2$\dotfill &RV Jitter Variance \dotfill &$17^{+38}_{-16}$\\
\enddata
\tablenotetext{}{See Table 3 in \citet{Eastman2019} for a detailed description of all parameters and all default (non-informative) priors.}
\tablenotetext{a}{Time of conjunction is commonly reported as the ``transit time.''}
\tablenotetext{b}{By the Lucy--Sweeney bias \citep{Lucy1971}, the reported eccentricity of the inner planet (Kepler-1514~c) is not significant. The orbit should be interpreted as consistent with circular.}
\tablenotetext{c}{Assumes no albedo and perfect redistribution.}
\end{deluxetable*}

\begin{figure}
    \centering
    \includegraphics[width=\columnwidth]{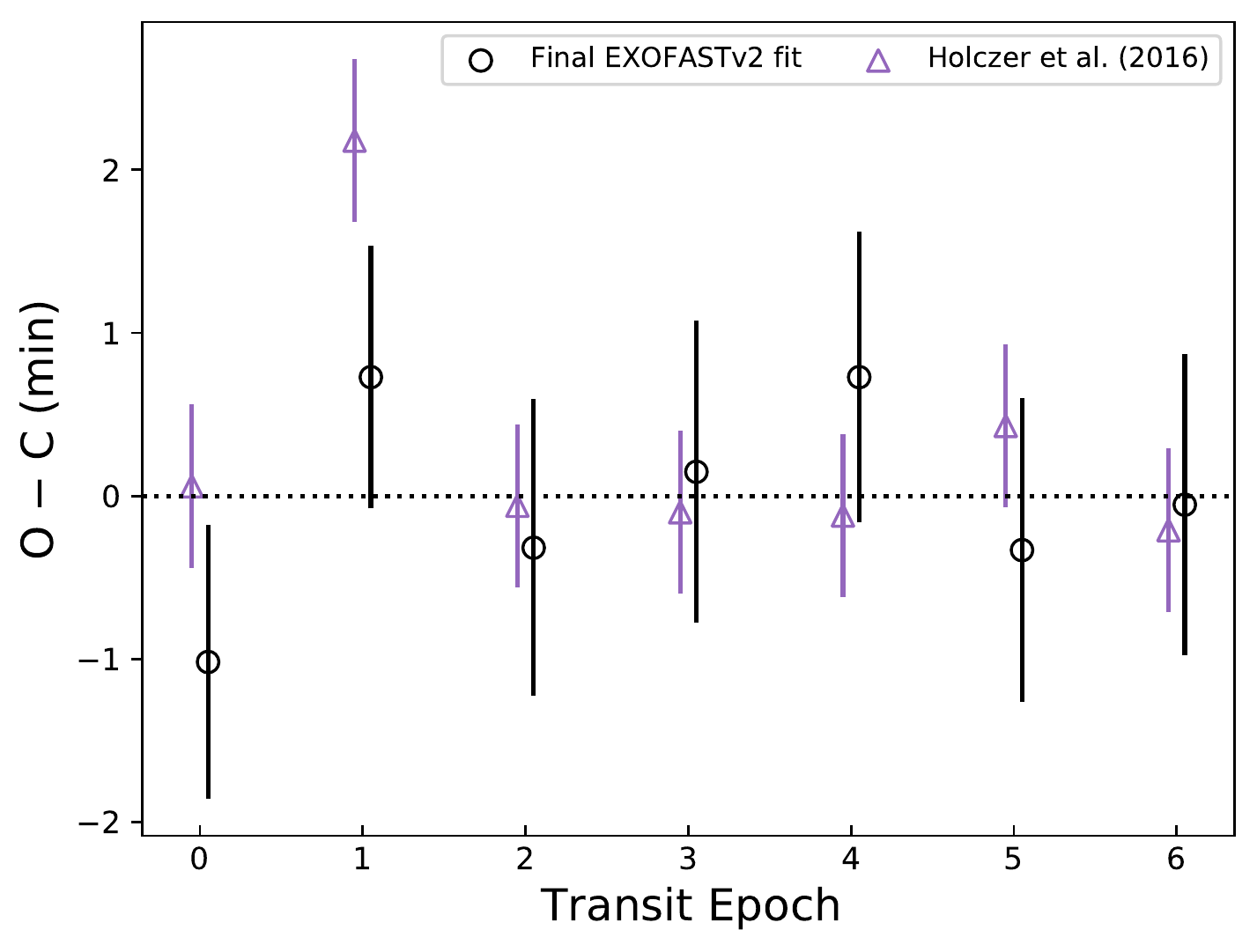}
    \caption{Observed minus calculated (O $-$ C) timing of the transits of \planet\ from the final, comprehensive \textsf{EXOFASTv2} fit (Section \ref{sec:final}). The measured times are broadly consistent with a linear ephemeris. The data sets have been offset horizontally for clarity.}
    \label{fig:ttvs2}
\end{figure}

The final TTVs for \planet\ are shown (as O$-$C values) in Figure \ref{fig:ttvs2}. As in the preliminary \textsf{EXOFASTv2} modeling, the statistical significance of the TTVs is weak. Although we cannot rule out dynamical interactions with other objects in the \host\ system as their source, their decreasing significance when incorporated into the system modeling indicates that they are likely the result of detrending and modeling choices related to stellar photometric variability. 

%%%%%%%%%%%%%%%%%%%%%%%%%%%%%%%%%%%%%%%%%%%%%%%%%%%%%%%%%%%%%%%%%%%%%%%%%%

\section{Results} 

\subsection{Confirming Kepler-1514~b}\label{sec:results}

\planet\ was originally deemed a planet through statistical validation by \citet{Morton2016}. Such validation for transiting exoplanets is fairly common, especially given how readily transiting exoplanets have been discovered. However, at orbital periods up to 400 days, suspected giant planet transit signals have an alarmingly high false positive probability \citep[e.g.,][]{Santerne2016}. Therefore, mass measurement is needed when confirming the planetary nature of a long-period ($P\gtrsim$100~days), giant exoplanet \citep[e.g.,][]{Dubber2019}.

We measure the mass of \planet\ to be {5.28$\pm$0.22~$M_{\rm J}$} and thereby confirm it to be a genuine planet. Its radius is {1.108$\pm$0.023~$R_{\rm J}$}, which places its bulk density in the 95th percentile among other weakly irradiated giant exoplanets. It orbits its host star with an orbital period of {217.83184$\pm$0.00012~days} and an orbital eccentricity of {0.401$^{+0.013}_{-0.014}$}. As we will discuss in the following sections, the combination of stellar, orbital, and planetary properties places it among a small group of interesting and accessible exoplanets.

%%%%%%%%%%%%%%%%%%%%%%%%%%%%%%%%%%%%%%%%%%%%%%%%%%%%%%%%%%%%%%%%%%%%%%%%%%

\subsection{Validating Kepler-1514~c}\label{sec:koi}

We did not infer the mass of \koi\ from the Keck-HIRES RVs in the final, comprehensive \textsf{EXOFASTv2} modeling because its signal is undetectable given the precision of the Keck-HIRES data (Section \ref{sec:final}). However, we are able to statistically validate the existence of this planet candidate.

We begin by ruling out the possibility of the transit signal originating from the neighbor star detected by \citet{Kraus2016}, which we determined is not associated with \host\ (see Section \ref{sec:imaging}). We follow the methodology of \citet{Vanderburg2019} to estimate the magnitude difference ($\Delta m$) between \host\ and the faintest possible neighbor that could cause the shallow transit signals. Equation 4 of \citet{Vanderburg2019} states 
\begin{equation}\label{eq:dmag}
    \Delta m \lesssim 2.5 \log_{10} \left ( \frac{t_{12}^2}{t_{13}^2 \delta} \right )
\end{equation}

\noindent where $t_{12}$ is the duration of transit ingress and egress (i.e., first to second contact), $t_{13}$ is the amount of time between first and third contact, and $\delta$ is the transit depth. The ingress and egress durations used in this calculation should not be constrained by stellar density, so we do not use results of the stellar modeling from Section \ref{sec:model}. Instead, we conduct a new fit to just the transits of \koi\ using \textsf{exoplanet}\footnote{\url{https://github.com/exoplanet-dev/exoplanet}} \citep{exoplanet}. This fit does not include any constraints based on stellar properties and all transit parameters are only bound to physically realistic regions of parameter space. We apply the same convergence criteria for this fit as for the \textsf{EXOFASTv2} fits described in Section \ref{sec:model}. After convergence, we derive values of $t_{12}$ and $t_{13}$ following Equations 14--16 of \citet{Winn2010}.

From Equation \ref{eq:dmag}, we find the distribution of $\Delta m$ values to be skewed toward zero, with median of 0.4 mag and a 99th percentile of 3.9 mag. We compare this value to the approximate \kepler-band magnitude of the neighbor star, which we estimate with a stellar population simulation from TRILEGAL \citep{Vanhollebeke2009,Girardi2005,Groenewegen2002} at the equatorial coordinates of \host. For simulated stars with $K_s$-band magnitudes of 16.7$\pm$0.5 (i.e., the sum of \host's magnitude and the NIRC2 imaging $\Delta m$), the distribution of \kepler-band magnitudes has a mean of 19.1 mag and a standard deviation of 0.8 mag. Compared with the \kepler-band magnitude of \host\ (11.69), this yields $\Delta m = 7.4 \pm 0.8$. The likely $\Delta m$ of the neighbor star in the \kepler-band is 8$\sigma$ discrepant with the median $\Delta m$ calculated in Equation 1, and over 4$\sigma$ discrepant with 99th percentile of the $\Delta m$ distribution. Therefore, we confidently rule out the neighbor star at a separation of 0$\farcs$27 as a possible cause of the \koi\ transits. 

\citet{Kraus2016} also reported the detection of three fainter neighbors ($\Delta K^{\prime} =$ 8.4--9.7) at wider separations 4$\farcs$1--5$\farcs$3. The \kepler-band $\Delta m$ values for these stars will be even larger than that of the close neighbor, so we can rule these stars out as the source of the \koi\ transits by the same argument.

Next, we use \textsf{VESPA} \citep{Morton2012,vespa2015} to calculate the false positive probability of \koi. We perform our calculation several times by drawing upon the inferred stellar properties and photometry of \host\ in addition to the contrast curve reported by \citet{Kraus2016}. In each calculation, the false positive probability was below the 1\% threshold typically employed for statistical validation.

The last piece of evidence we provide for the validation of \koi\ is the results of \cite{Lissauer2012}, which show that a vast majority of \kepler\ multi-planet candidates are indeed genuine planets. Specifically, the study estimates that in systems with 1 confirmed planet and 1 planet candidate, the planet candidate is a false positive $<1\%$ of the time. This combination of this information and that provided above makes a thorough case for the validation of this planet candidate. Therefore, based on our validation analysis, we hereafter refer to \koi\ as Kepler-1514~c.

%%%%%%%%%%%%%%%%%%%%%%%%%%%%%%%%%%%%%%%%%%%%%%%%%%%%%%%%%%%%%%%%%%%%%%%%%%

\section{Discussion}\label{sec:disc}

\subsection{Tension in Stellar Properties}

The stellar properties of the \host\ system are constrained by both the SED data and the transit and RV data included in the comprehensive modeling (Section \ref{sec:final}). We explored how each of these affected the final stellar properties (Table \ref{tab:stellar}) by running two additional \textsf{EXOFASTv2} fits. The first was a ``star only'' fit (i.e., with no transit or RV data), and the second was a ``no SED'' fit (i.e., identical to the global fit but without the SED). In lieu of the SED, we applied a prior to stellar effective temperature (6073$\pm$110~K) based on spectroscopic analysis of the high S/N template spectrum. In the ``star only'' fit, \host\ was found to be more massive ($M_{\star} = 1.252^{+0.050}_{-0.064}$~$M_{\sun}$), denser ($\rho_{\star} = 0.918^{+0.080}_{-0.095}$~g~cm$^{-3}$), and hotter ($T_{\rm eff} = 6470\pm170$~K) when compared to the same parameters in the ``no SED'' fit ($M_{\star} = 1.102^{+0.089}_{-0.087}$~$M_{\sun}$; $\rho_{\star} = 0.783^{+0.046}_{-0.044}$~g~cm$^{-3}$; $T_{\rm eff} = 5982^{+93}_{-87}$~K). The stellar radii inferred from these two fits were consistent, but in mass, density, and effective temperature, the discrepancies were $1.4\sigma$, $1.3\sigma$, and $2.5\sigma$, respectively. Our final solution, as presented in Section~\ref{sec:final}, represents a compromise between these two slightly discrepant solutions, though it is likely that our uncertainties are slightly underestimated. Although this tension is passed down to the planetary parameters as well, it does not affect our interpretation of the planets themselves.

This slight tension is due to a mismatch between the stellar mass and radius from the MIST models and SED, respectively, and the stellar density constrained by the transit duration and eccentricity \citep{Seager2003}. It is unclear which to believe more. On one hand, the transits have a very high S/N but half of the RV phase curve is sparsely sampled (i.e., there are only two data points between $-0.5$ and 0 in Figure \ref{fig:rv}). If the eccentricity were biased high by either of these points, it would skew the inferred stellar density and could be the source of this tension. While we see no evidence to suggest either point is problematic, many undetectable problems could lead to significant single point RV outliers. On the other hand, the stellar models that underlie the MIST and SED constraints have poorly understood systematics. \textsf{EXOFASTv2} automatically attempts to account for them, but it may not be sufficient.

One way to further investigate this tension is to acquire more high precision RV observations that cover the sparsely sampled phases. Ideally, this would eliminate the possibility that the stellar density is being influenced by a single point outlier in the RV data set. The transit and RV data for most exoplanet systems are not precise enough to produce a constraint on stellar density that can overwhelm the stellar information present in the isochrone models and SED, especially when precise \gaia\ parallax measurements are used. In this way, the \host\ system could provide valuable future tests of stellar models that otherwise limit measurements of fundamental stellar properties \citep{Tayar2020}.

%%%%%%%%%%%%%%%%%%%%%%%%%%%%%%%%%%%%%%%%%%%%%%%%%%%%%%%%%%%%%%%%%%%%%%%%%%

\subsection{Kepler-1514~b: A Dense, Cool Giant Planet}\label{sec:disc1}

When considering \planet\ among other known exoplanets, the foremost point of interest is its transiting geometry despite it 218-d orbit. This property places \planet\ in the 98th percentile of transiting exoplanets by orbital period. Considering the planet characterization opportunities enabled by transits, \planet\ is in an inherently interesting group of exoplanets.  

With a longer orbital period also comes a lower stellar irradiation relative to most transiting exoplanets. \planet\ receives an average incident flux of {4.4\e{6}~erg~s$^{-1}$~cm$^{-2}$ (3.2 times that of Earth)}, which is approximately two orders of magnitude below the empirically determined threshold for radius inflation \citep{Miller2011,Demory2011b}. \planet\ is still informative to investigations of radius inflation, though. \citet{Sestovic2018} found that giant planet radius inflation is a function of planet mass, and for giant planets with $M_p>2.5M_{\rm J}$, radius inflation is not effective below $\sim$1.6\e{8}~erg~s$^{-1}$~cm$^{-2}$ incident flux. However, the weakly irradiated side of this threshold for massive giant planets contains only two planets. Adding \planet\ as a third member to this small group would likely inform the radius inflation boundary for massive planets. 

\begin{figure}
  \begin{center}
    \begin{tabular}{cccc}
      \includegraphics[width=\columnwidth]{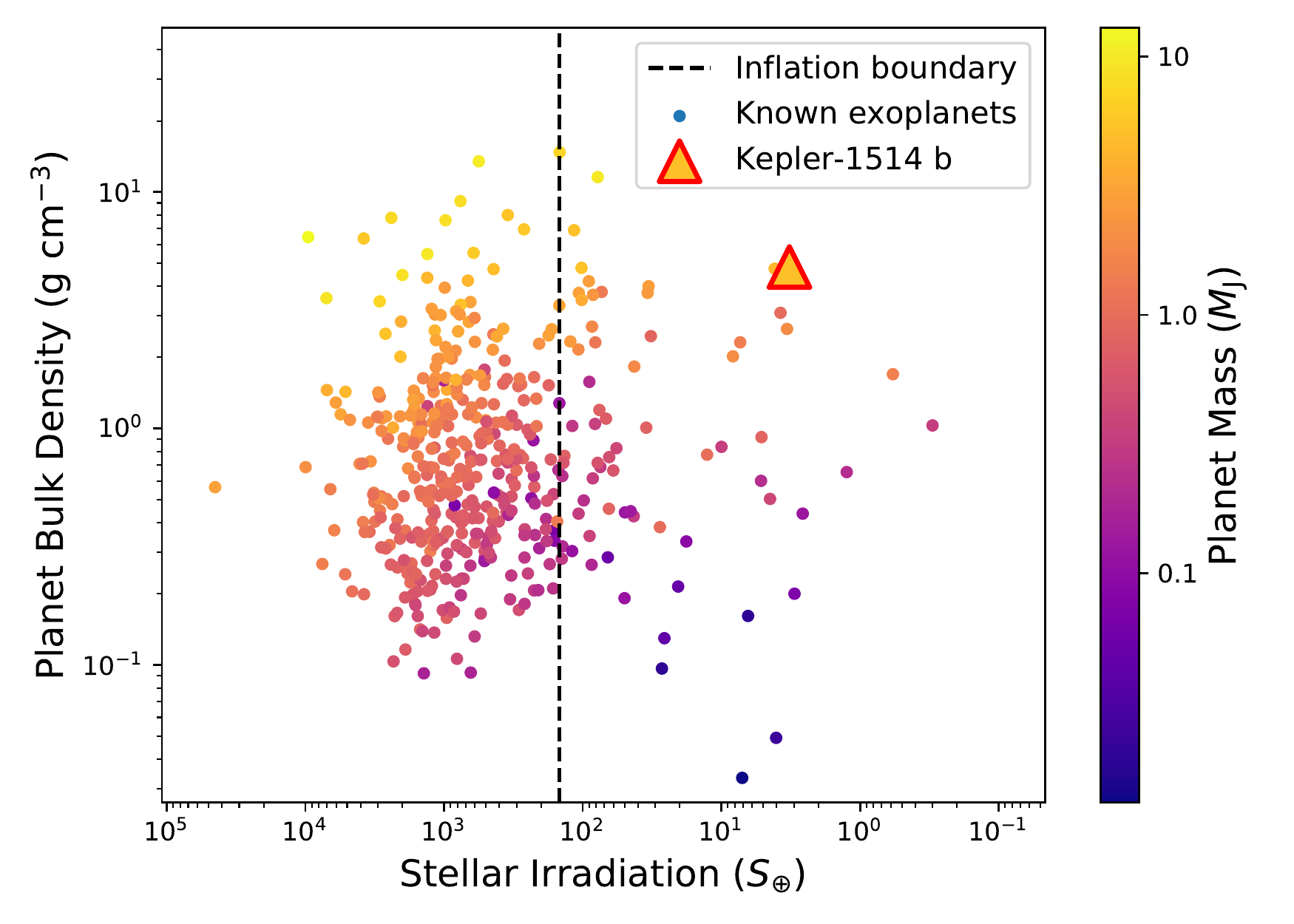} \\
      \includegraphics[width=\columnwidth]{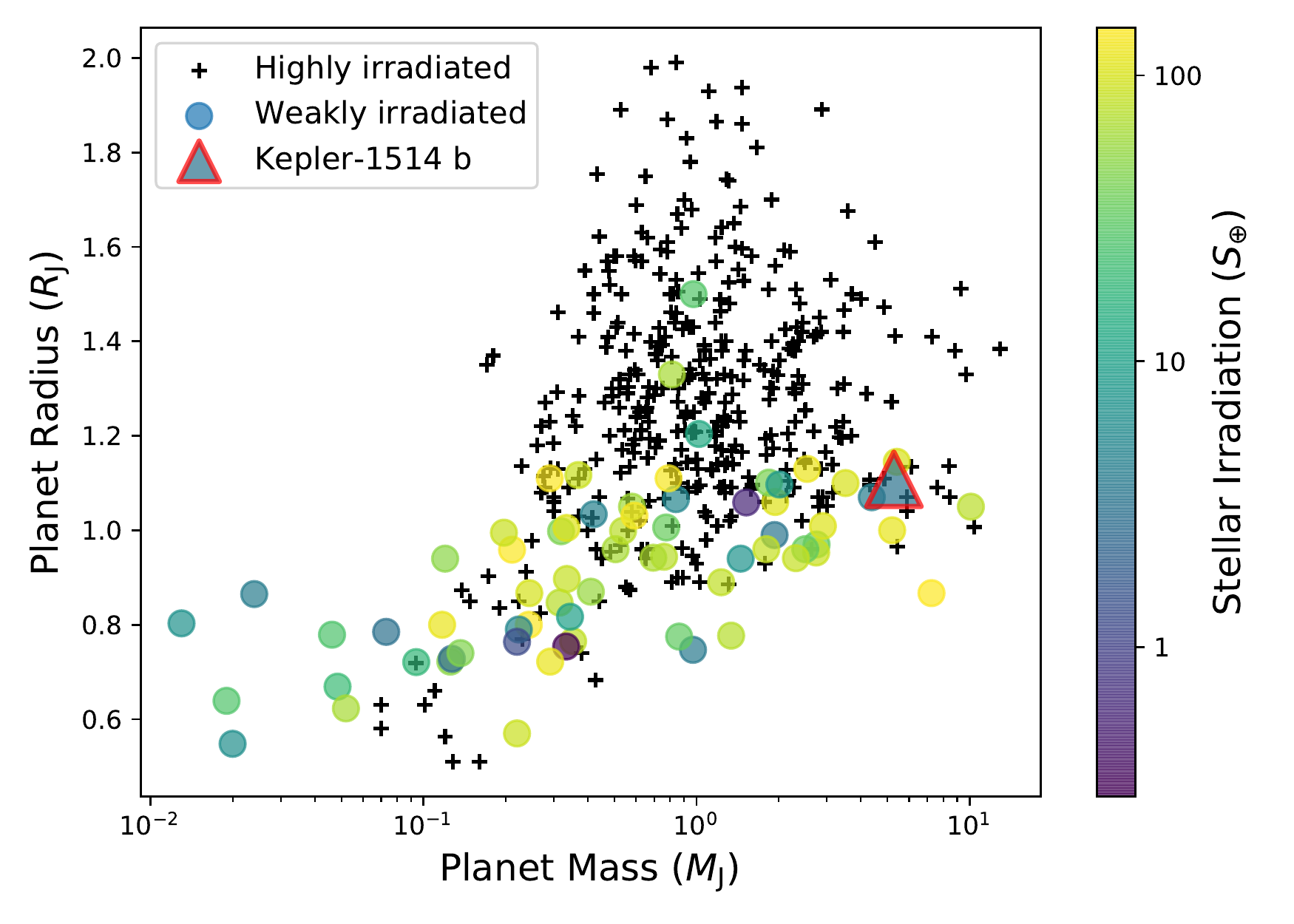} 
    \end{tabular}
  \end{center}
  \caption{All confirmed giant ($R_p>$ 0.5~$R_{\rm J}$) exoplanets (from the NASA Exoplanet Archive; accessed 2020 July 9) for which stellar irradiation was either given or could be calculated and planet mass and radius were known to at least 50\% precision. {\it Top:} of those planets with stellar irradiation below the empirical inflation boundary \citep{Miller2011,Demory2011b}, \planet\ ranks in the 95th percentile in bulk density. The spread in density is due to the spread in mass, since most of these weakly irradiated giant planets are roughly the same size. {\it Bottom:} the inflation boundary from the top panel separates weakly and highly irradiated planets. The combination of high mass and low irradiation for \planet\ places it among a small group of giant planets that are useful for testing models of giant planet interior structure.}
  \label{fig:rho_vs_S}
\end{figure}

The Jupiter-sized \planet\ has a bulk density of {4.82$^{+0.26}_{-0.25}$~g~cm$^{-3}$}, which is consistent with that of other cold, giant planets for which electron degeneracy pressure yields high densities \citep[e.g.,][]{Weiss2013}. Among other known giant planets receiving flux below the canonical radius inflation threshold, \planet\ ranks in the 95th percentile by bulk density (Figure \ref{fig:rho_vs_S}, top panel). It marks the upper tail of a distribution of bulk density that spans two orders of magnitude, mirroring a similar spread in planet mass (as indicated by colors of the points in Figure \ref{fig:rho_vs_S}). 

In mass-radius space (Figure \ref{fig:rho_vs_S}, bottom panel), \planet\ occupies a region where planet size has become almost entirely independent of mass. Different studies have suggested a range of masses at which electron degeneracy pressure becomes the primary source of support within a giant planet's interior, leading to  increasingly more massive objects of nearly the same size. The early theoretical work by \citet{Zapolsky1969} found this mass to be between 1.2 and 3.3 $M_{\rm J}$ for an isolated sphere of hydrogen and helium. More recent planetary evolution models \citep{Fortney2007} suggest a range of roughly 2--5~$M_{\rm J}$ depending on composition and stellar irradiation. Empirical measurements of the transition to degenerate cores have included $\sim$0.5~$M_{\rm J}$ \citep{Weiss2013} and 0.41$\pm$0.07~$M_{\rm J}$ \citep{Chen2017}. The former value was a fiducial boundary that represents a broad peak extending up to several Jupiter masses \citep[see Figure 12 of][]{Weiss2013}, while the latter value was inferred from data without assuming prior knowledge of giant planet structure. In either case, the discrepancy with the previously mentioned models may, at least in part, be due to planetary radii that are inflated by physical mechanisms not captured by the models. Nevertheless, at {5.3~$M_{\rm J}$}, \planet\ is likely supported through electron degeneracy pressure. Considering only the weakly irradiated giant planets in Figure \ref{fig:rho_vs_S} (bottom panel), only a few have masses as large as or greater than \planet. These planet are valuable laboratories for testing models of models of giant planet interiors. \planet\ specifically adds a crucial new data point at high density and low insolation that is especially amenable to explorations of interior metallicity and evolution.

In mass, radius, density, and average stellar irradiation, \planet\ is similar to HD~80606~b \citep[$M_p \approx 4.1$~$M_{\rm J}$, $R_p \approx 1.0$~$R_{\rm J}$, $\rho_p \approx 5.1$~g~cm$^{-3}$, and $S_p \approx 4.1$~$S_{\earth}$;][]{Bonomo2017}. The orbit of \planet\ is also moderately eccentric, although substantially less than that of HD~80606~b \citep[$e\approx0.93$;][]{Bonomo2017}. Despite these similarities, their formation histories may be different. The high eccentricity of HD~80606~b is thought to be a remnant of migration driven by an associated stellar companion \citep[e.g.,][]{Naef2001,Moutou2009}. As discussed in Section \ref{sec:imaging}, the only known nearby neighbor of \host\ is a background source. Combined with the semi-major axis and eccentricity of \planet's orbit and the stellar metallicity (i.e., [Fe/H]), \planet\ may have instead migrated via planet-planet scattering \citep[e.g.,][]{Dawson2018} or within a cavity formed in the protostellar disk, the latter of which is perhaps more consistent with the presence of Kepler-1514~c \citep[e.g.,][]{Debras2021}. All of the other similarities between \planet\ and HD~80606~b are interesting to consider in light of possible different migration pathways. Further data, and possibly numerical simulations that include the inner planet Kepler-1514~c, would be useful to place stronger constraints on evolutionary theories.

%%%%%%%%%%%%%%%%%%%%%%%%%%%%%%%%%%%%%%%%%%%%%%%%%%%%%%%%%%%%%%%%%%%%%%%%%%

\subsection{Further Study: Interiors, Atmospheres, Obliquity, and Exomoons}\label{sec:disc2}

One avenue of continued study is to consider the interior structure of the giant planet \planet. \citet{Thorngren2016} identified a relationship between increasing mass and increasing heavy element mass for uninflated giant exoplanets. However, for planet mass greater than $\sim$3~$M_{\rm J}$, this relationship was informed by only three data points that showed substantial scatter \citep[see Figure 11 of][]{Thorngren2016}. Furthermore, \citet{Thorngren2016} also identified an inverse relationship between planet mass and metal enrichment relative to stellar for the same sample of weakly irradiated giant planets. As found by the spectroscopic stellar characterization (Section \ref{sec:final}), \host\ is only slightly metal-rich ({[Fe/H] = 0.119$^{+0.080}_{-0.075}$}, Table \ref{tab:stellar}). Testing for a weak relative metal enhancement between \planet\ and its host through a metallicity retrieval or an atmospheric abundance measurement would be helpful to refining both aforementioned relationships. 

\begin{figure}
    \centering
    \includegraphics[width=\columnwidth]{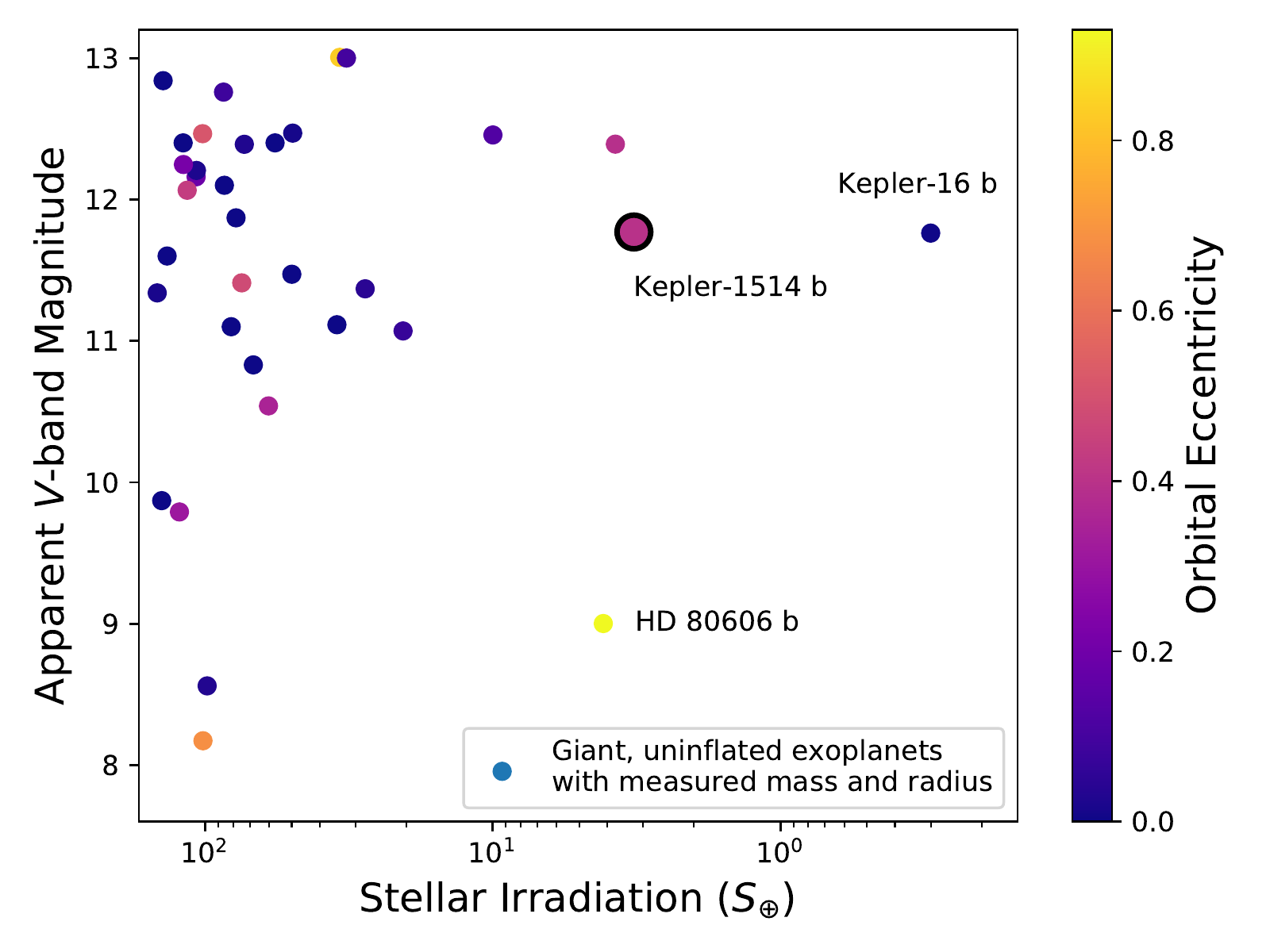}
    \caption{Giant ($R_p>0.5R_{\rm J}$) exoplanets with mass and radius measured to better than 50\% precision that receive stellar irradiation below 2\e{8}~erg~s$^{-1}$~cm$^{-2}$ stellar, meaning they are likely uninflated \citep[e.g.,][]{Miller2011,Demory2011b,Sestovic2018}. The points are colored by orbital eccentricity (gray if not reported).}
    \label{fig:mag_vs_S}
\end{figure}

A key aspect of the amenability of the \host\ system to the follow-up characterization we have discussed here is the stellar brightness. \host\ has a $V$-band magnitude of 11.8. Of all the planet host stars discovered by the \kepler\ primary mission, only 81 are brighter at optical wavelengths. This brightness is especially valuable when comparing to other weakly irradiated giant exoplanet systems with known masses and radii (Figure \ref{fig:mag_vs_S}). At similar brightness, only Kepler-16~b receives a lower stellar irradiation. At similar stellar irradiation, only HD~80606~b is brighter. Together, these three exoplanets are representative of broad diversity in orbital eccentricities of long-period giant planets as well.  

Despite the promising brightness of \host, prospects for atmospheric characterization via transmission spectroscopy are poor. The high mass of the \planet\ yields a surface gravity of {$\sim$107~m~s$^{-2}$}, much higher than that of Jupiter ($\sim$25~m\,s$^{-2}$) or Saturn ($\sim$10~m\,s$^{-2}$). Adopting the equilibrium temperature (Table \ref{tab:planets}) and assuming a hydrogen dominated atmosphere, we estimate an atmospheric scale height of {$\sim$15~km}. A transmission spectrum feature of a few scale heights would only be {$\sim$10 parts per million}, even in the absence of clouds, which is beyond the reach of any current or planned observational facility. Similarly, atmospheric characterization via direct imaging is also challenging, as the separation between \planet\ and its host star is only 2~mas. 

Another exciting avenue of further study of \planet\ is the measurement of stellar obliquity through the Rossiter-McLaughlin (RM) effect \citep{Rossiter1924,McLaughlin1924}. Spin-orbit alignment plays a key role in planetary migration processes \citep[e.g.,][]{Fabrycky2007,Chatterjee2008}, so determining this value for \planet\ would be particularly revealing. Using the high S/N template spectrum of \host\ acquired with Keck-HIRES (see Section \ref{sec:hires}), we measured the stellar projected rotational velocity ($v\sin{i}$) to be $7.8\pm1.0$~km~s$^{-1}$ following the spectral matching technique of \citet{Petigura2017a}. According to Equation 40 of \citet{Winn2010}, we would therefore expect the amplitude of the RM effect to be {$\sim$60~m~s$^{-1}$}. The 21 hr transit duration presents a formidable challenge, though, as it is longer than the maximum length of time that any single site with precise RV capabilities can observe the star. Depending on the transit timing and the precision of the RV facility, it may be possible to detect the RM effect in an observation of a partial transit (i.e., baseline and ingress or egress). The Keck-HIRES observations of \host\ achieved $\sim$5~m~s$^{-1}$ internal precision with exposure times between 10 and 19 minutes (depending on observing conditions). Assuming stable 15-minute exposures, we could acquire $\sim$7 RV measurements with $\sim$5~m~s$^{-1}$ uncertainty over the 1.78~hr ingress (or egress) with Keck-HIRES. This may be sufficient to constrain the stellar obliquity. Alternatively, the \host\ system may be an opportunity for a coordinated observing campaign at multiple sites spread out in longitude assuming that the noise properties of both facilities are well characterized. In either case, further effort should be made to explore the extent to which RM measurements of partial transits of long-period exoplanets lead to degeneracies in the solutions for stellar obliquity.

To date, the majority of systems subject to RM measurements host short-period hot Jupiters \citep[see][for a review]{Triaud2018}. Currently, Kepler-16 is the only system with stellar obliquity measurement from a planet with a longer orbital period ($P=228$~days) than \planet\ \citep{Winn2011}. However, Kepler-16 is a binary system. This means that \planet\ is poised to become the longest-period exoplanet with a stellar obliquity measurement in a single star system. 

Lastly, we point out the potential of \planet\ as a host for exomoons. It is plausible that a massive, giant planet with an orbital period of several hundred days may harbor a system of exomoons.  \citet{Teachey2018a} estimated the occurrence of Galilean-size exomoons for exoplanets similar to \planet\ to be $0.16^{+0.13}_{-0.10}$. \citet{Hill2018} also discussed the occurrence of exomoons orbiting long-period giant planets discovered by \kepler, suggesting the possible existence of a large population of exomoons within their star's habitable zones. Furthermore, several other efforts to identify exomoons have recognized \planet\ \citep{Kipping2012,Kipping2015,Guimaraes2018}. We demonstrated that \planet\ exhibits weak TTVs (Section \ref{sec:model}), which could have several explanations including exomoons \citep[e.g.,][]{Sartoretti1999,Szabo2006,Simon2007,Kipping2009a,Kipping2009b}. 

However, we presently do not have evidence to support such an extraordinary claim. Relative to the Solar System giant planets---that are known to host moons in abundance---\planet\ likely experienced a different formation and migration history that may have involved processes that are thought to deplete planets of moons \citep[e.g.,][]{Barnes2002,Spalding2016}. Recent large scale efforts have broadly applied new techniques to identify exomoon host candidates in data from transit surveys including \kepler\ \citep{Kipping2020,Rodenbeck2020}. Now that the long-period giant planet \planet\ has had its mass measured, the \host\ system is likely worth revisiting for a more focused investigation on the possible existence and detectability of exomoon candidates.

%%%%%%%%%%%%%%%%%%%%%%%%%%%%%%%%%%%%%%%%%%%%%%%%%%%%%%%%%%%%%%%%%%%%%%%%%%

\section{Summary}\label{sec:conc}

We conducted RV observations of \host\ using the HIRES instrument on the Keck I telescope. Based on data collected by the primary \kepler\ mission \citep{Borucki2010} and analysis conducted by \citet{Morton2016}, this system was thought to contain a cool gas giant planet on a 218~d orbital period (that was statistically validated) and a shorter-period Earth-size KOI. The transits of each object in the \host\ system displayed variations in timing (relative to a linear ephemeris), depth, and duration \citep{Holczer2016}. Inspired by the high false positive probability of long-period ($P\gtrsim$100~days), giant planet signals in \kepler\ transit data \citep{Santerne2016} and also by the inherent rarity of long-period transiting exoplanets, we aim to measure the mass of \planet\ and characterize the system.

We apply spline detrending to remove the stellar variability of the host star present in the \kepler\ photometry (Section \ref{sec:spline}). This detrending casts doubt upon a dynamical explanation for the TTVs and T$\delta$Vs (see Section \ref{sec:model}) but we nonetheless include the former in the comprehensive global modeling of the transit and RV data. The RV observations (Section \ref{sec:hires}) readily identify a planetary, Keplerian signal corresponding to \planet, which we find to be massive ({$M_p = 5.28\pm0.22$~$M_{\rm J}$}) and on a moderately eccentric orbit ({$e=0.401^{+0.013}_{-0.014}$}). The modest set of RVs, although precise, is not able to constrain the mass of \koi, for which we expect a sub-meter-per-second RV semi-amplitude. However, through a false positive probability analysis that includes scenarios introduced by neighboring stars, we validate the planetary nature of \koi\ (now known as Kepler-1514~c) with a false-positive probability below 1\% (Section \ref{sec:koi}). 

Based on these results, we postulate on the possible interior properties and formation history of \planet\ and its utility as one of only a select few long-period ($P>$100~days) giant exoplanets with a well known mass \textit{and} radius (Section \ref{sec:disc1}). \planet\ is unlikely to be inflated \citep[e.g.,][]{Miller2011,Demory2011b} like its hot Jupiter counterparts, but its relatively high mass makes it a useful test of the radius inflation thresholds put forth by \citet{Sestovic2018}. Based on the lack of a known associated stellar companion (Section \ref{sec:imaging}), we assert that \planet\ may have migrated via planet-planet scattering, although we cannot rule out other mechanisms. The high bulk density \planet\ ({$4.82^{+0.26}_{-0.25}$~g~cm$^{-3}$}) is atypical among giant planets, but is consistent with those having nearly constant radius above $\sim$0.5~$M_{\rm J}$ masses because of electron degeneracy pressure.

Moving forward, we consider \planet\ as a candidate for further investigation (Section \ref{sec:disc2}). Although prospects for atmospheric characterization via transmission spectroscopy are poor, the system is highly amenable to a stellar obliquity measurement via the RM effect. Furthermore, \planet\ has been previously identified as a promising system for searches for exomoons. With the new mass measurement presented here, we recommend a focused reexamination of the \host\ system and its potential to harbor natural satellites.   

We note that, during the preparation of this manuscript, \koi\ was statistically validated as Kepler-1514~c by \citet{Armstrong2020}.

%%%%%%%%%%%%%%%%%%%%%%%%%%%%%%%%%%%%%%%%%%%%%%%%%%%%%%%%%%%%%%%%%%%%%%%%%%

\acknowledgements

The authors thank the anonymous referee for thoughtful comments that improved the quality and clarity of this work. The authors thank all of the observers in the California Planet Search team for their many hours of hard work. P. D. is supported by a National Science Foundation (NSF) Astronomy and Astrophysics Postdoctoral Fellowship under award AST-1903811. This research has made use of the NASA Exoplanet Archive, which is operated by the California Institute of Technology, under contract with the National Aeronautics and Space Administration under the Exoplanet Exploration Program. This research made use of \textsf{exoplanet} and its dependencies \citep{Kipping2013b,astropy2013,astropy2018,Luger2018,Agol2020,pymc3,theano}. 

This paper includes data collected by the \kepler\ mission and obtained from the MAST data archive at the Space Telescope Science Institute (STScI). Funding for the \kepler\ mission is provided by the NASA Science Mission Directorate. STScI is operated by the Association of Universities for Research in Astronomy, Inc., under NASA contract NAS 5–26555. Some of the data presented herein were obtained at the W. M. Keck Observatory, which is operated as a scientific partnership among the California Institute of Technology, the University of California, and NASA. The Observatory was made possible by the generous financial support of the W.M. Keck Foundation. Finally, the authors wish to recognize and acknowledge the very significant cultural role and reverence that the summit of Maunakea has always had within the indigenous Hawaiian community. We are most fortunate to have the opportunity to conduct observations from this mountain.

\facilities{Keck:I (HIRES), Keck:II (NIRC2), Kepler}

\software{   \textsf{astropy} \citep{astropy2013,astropy2018},
                \textsf{EXOFASTv2} \citep{Eastman2013,Eastman2017,Eastman2019}, 
                \textsf{VESPA} \citep{Morton2012,vespa2015},
                \textsf{exoplanet} \citep{exoplanet},
                \textsf{pymc3} \citep{pymc3},
                \textsf{theano}, \citep{theano}
                }

\end{document}